\documentstyle{article}
\begin{document}
\large
\baselineskip=24pt
\title{Stability Analysis of
\\Lattice Boltzmann Methods\\[.2cm]}
\author{James D. Sterling \\
  [.2cm]{\em Center for Nonlinear Studies, MS-B258}\\
	{\em Los Alamos National Laboratory}\\
	{\em Los Alamos, NM 87545}\\[.2cm]
{\em Permanent Address: Advanced Projects Research Incorporated}\\ {\em 5301 N.
Commerce Ave., Suite A, Moorpark, CA 93021}\\[.7cm]
        {Shiyi Chen}\\
  [.2cm]{\em Theoretical Division, MS-B213}\\
	{\em Los Alamos National Laboratory}\\
	{\em Los Alamos, NM 87545}}
\date{}
\maketitle
\newpage
\begin{center}
\large{\bf ABSTRACT}
\end{center}
\baselineskip=12pt
The lattice Boltzmann equation describes the evolution of the velocity
distribution function on a lattice in a manner that macroscopic fluid dynamical
behavior is recovered. Although the equation is a derivative of lattice gas
automata, it may be interpreted as a Lagrangian finite-difference method for
the numerical simulation of the discrete-velocity Boltzmann equation that makes
use of a BGK collision operator. As a result, it is not surprising that
numerical instability of lattice Boltzmann methods have been frequently
encountered by researchers. We present an analysis of the stability of
perturbations of the particle populations linearized about equilibrium values
corresponding to a constant-density uniform mean flow. The linear stability
depends on the following parameters: the distribution of the mass at a site
between the different discrete speeds, the BGK relaxation time, the mean
velocity, and the wavenumber of the perturbations. This parameter space is too
large to compute the complete stability characteristics. We report some
stability results for a subset of the parameter space for a 7-velocity
hexagonal lattice, a 9-velocity square lattice and a 15-velocity cubic lattice.
Results common to all three lattices are 1) the BGK relaxation time $\tau$ must
be greater than $\frac{1}{2}$ corresponding to positive shear viscosity, 2)
there exists a maximum stable mean velocity for fixed values of the other
parameters and 3) as $\tau$ is increased from $\frac{1}{2}$ the maximum stable
velocity increases monotonically until some fixed velocity is reached which
does not change for larger $\tau$.

\section{Introduction}
The lattice Boltzmann (LB) method is a recently developed computational scheme
used to model fluids under a variety of flow regimes. As a derivative of
lattice gas (LG) automata, the LB method deals with fluid dynamics from the
microscopic, kinetic level. However, as with the Boltzmann equation, the LB
method describes the evolution of particle populations rather than attempting
to follow individual particle motion. Thus, the LB method has the flexibility
of traditional particle methods  but has the numerical character of
finite-difference schemes. The physical interpretation of the scheme as
consisting of a particle streaming step followed by a collision results in a
very simple parallel logic that is well suited for implementation on massively
parallel computers. The main advantage of the LB method is that the particle
interpretation allows the use of very simple boundary conditions so that the
parallel implementation may be used even for complex geometries. For this
reason, one of the most successful applications of the LB method has been to
simulations of flow through porous media \cite{roth1} \cite{chen}.

The development of LG models was based on the observation that macroscopic
behavior of fluid flow is not very sensitive to the underlying microscopic
physics. Thus, models were developed based on the simplest possible particle
microworld that would lead to the incompressible Navier-Stokes equation in the
limit of small Knudsen number \cite{FHP}. The methods successfully modeled
incompressible fluid flow but noise associated with the particle microworld
necessitated the introduction of some type of averaging procedure such as
spatial, temporal or ensemble averaging to characterize the macroscopic flow. A
second difficulty is that LG methods have unphysical equations of state and
non-Galilean invariant flow. Finally, the transport coefficients that resulted
from the microscopic collision rules were inflexibly limited to small ranges of
values \cite{Diemer}.

In contrast, the numerical solution of the lattice Boltzmann equation (LBE), as
proposed by McNamara and Zanetti \cite{mcn}, neglects individual particle
motion resulting in smooth macroscopic behavior. Further simplification of the
scheme is achieved by linearizing the collision operator \cite{higuera}. A
particularly simple linearized version of the collision operator makes use of a
relaxation towards an equilibrium value using a single relaxation time
parameter. The relaxation term is known as the BGK \cite{BGK} collision
operator and has been independently suggested by several authors for use with
this method \cite{koelman} \cite{chent} \cite{qian}. Use of this collsion
operator makes the computations much faster and allows flexibility of the
transport coefficients.  Particle streaming and collision are explicitly
computed by performing a type of ``shift" operation on the parallel computer to
represent the particle streaming followed by a purely local operation for the
collision. The microscopic approach of the LB method associates physical
quantities with the discretization parameters: the time step is the time
between particle collisions and the lattice spacing is proportional to the mean
free path. Again, the spirit of the approach is to retain the simplest
microscopic description that gives the macroscopic behavior of interest.

Application of a Taylor series expansion of the lattice kinetic equation
followed by a Chapman-Enskog expansion results in the typical hierarchy of
equations; Euler, Navier-Stokes, Burnett, etc.  By selecting the appropriate
number of speeds and the appropriate form of the equilibrium distribution
function, one may match the equations that result from the LB method with those
of traditional kinetic theory to the desired level. Higher level terms that are
not matched represent behavior of the lattice gas that differs from a
Maxwellian gas.

The most common application of the LB method has been to fluid flow models for
which only mass and momentum are conserved. The Chapman-Enskog theory for these
models typically yields correct behavior to the Euler level but the
Navier-Stokes level is correct only in the incompressible limit. In other
words, the incorrect terms become small as the square of the Mach number
becomes small. This approach has much in common with explicit ``penalty" or
``pseudocompressibility" methods of solving incompressible flows \cite{chorin}
\cite{dukowicz} \cite{ramshaw}. Complete energy-conserving models that yield
the correct form of the compressible continuity, momentum, and energy equations
have been developed by Alexander, Chen, and Sterling \cite{alex} and by
McNamara and Alder \cite{guy}. We note that for any of the LB models, the
transport coefficients depend on the time step and lattice spacing. Another way
of looking at this is that there is a ``lattice viscosity" or ``numerical
viscosity" that becomes small as the grid is refined (i.e. time step reduced
for fixed particle velocities). This brings us to an alternative view that the
higher order terms in the Taylor series expansion of the kinetic equation are
not ``physical" but may be considered ``truncation error" of a finite
difference approximation to some continuous equation.

Indeed, an alternative view of the LB method is that it is a particular space
and time discretization of the discrete-velocity Boltzmann equations. These
equations are partial differential equations (i.e. continuous in space and
time) that describe the evolution of particle populations that have discrete
speeds. Researchers have used a variety of discrete-velocity models: models
with a single speed were originally developed by Broadwell \cite{broadwell} and
recent work by Inamuro and Sturtevant \cite{Inamuro} includes many speeds.
Ancona \cite{mario} introduces the view that the LB method is a
finite-difference method for the solution of the macroscopic equations and
generalizes the method to include fully-Lagrangian methods for the solution of
partial differential equations. Other higher-order finite difference methods
may also be used to approximate the discrete-velocity Boltzmann equation. These
alternative discretizations result in new ``lattice-Boltzmann equations" that
may be, in turn, viewed as microscopic models with particular streaming and
collision processes. A Taylor series expansion of the resulting higher-order
lattice Boltzmann equations will not have discretization parameters that enter
at the Navier-Stokes level of the Chapman-Enskog procedure. Thus, the transport
coeffients at this level will not have any associated ``lattice" contribution
and convergence to the Navier-Stokes equations will be of higher order by
design.

We note that the Chapman-Enskog procedure makes use of expansions in a small
parameter which is proportional to the lattice spacing of the LB method.
However, by the nature of asymptotic expansions, the resulting behavior is
typically not very sensitive to the size of the small parameter. Thus, the
accuracy of the finite-difference approximation to the discrete Boltzmann
equation can be maximized by choosing small discretization parameters. Inamuro
and Sturtevant \cite{Inamuro} used first, second and third order upwind finite
difference discretizations of the discrete-velocity Boltzmann equation to study
shock-wave structure, conductive heat transfer, and chemical vapor deposition.
However, they made use of a large velocity set that was nearly Maxwellian in
distribution and since their intention was to model rarefied flows, no
Chapman-Enskog procedure was used to assess continuum-limit behavior. In
contrast, Reider and Sterling \cite{reider} have studied the convergence
behavior of different finite-difference approximations to the
discrete-Boltzmann equation for velocity sets that provide Navier-Stokes
behavior in the incompressible and continuum limits. We will not address
numerical accuracy in this paper but will be concerned with another aspect of
numerical analysis for the lattice-Boltzmann method: stability.

In traditional kinetic theory, the equilibrium velocity distribution function
is the maximum entropy state. Thus, any initial state will evolve towards a
state of higher entropy. This result is known as Boltzmann's H-theorem which
ensures an increase of entropy, and ensures stability. An H-theorem has been
derived for some particle methods and a derivation for Lattice Gases is
included in reference \cite{FHP2}. If one can guarantee that the equilibrium
distribution function for LB methods is the maximum entropy state, then
stability can be guaranteed even though LB approaches are not particle methods
\cite{Levermore}. The problem with this approach however, is that one cannot
usually find an equilibrium distribution function that can simultaneously
guarantee an H-theorem and allow the correct form of the equations to be
obtained. In this paper, we limit our discussion to LB schemes that have been
developed for simulating the incompressible Navier-Stokes equations (i.e.
simulation in the low Knudsen number and low Mach number limits). These schemes
do not have an H-theorem and are therefore subject to numerical instability.
Lattice Boltzmann results that are reported in the literature have typically
been performed under conditions that provide stable behavior. However, it is
well known among LB researchers that instability problems arise frequently.
When the LB method is viewed as a finite-difference method for solving the
discrete-velocity Boltzmann equations, it becomes clear that numerical accuracy
and stability issues should be addressed.

In section 2 a review of the Chapman-Enskog procedure is provided and the
method is applied to derive the macroscopic equations corresponding to a
7-velocity hexagonal lattice model. Section 3 presents the lattice Boltzmann
discretization and provides an assessment of the order of this numerical
scheme. In addition, the simple physical interpretation of the lattice
Boltzmann equation and its implications for boundary conditions are discussed.
In Section 4 we present a von Neumann stability analysis of the lattice
Boltzmann method for a uniform flow and report results for a 7-velocity
hexagonal lattice, a 9-velocity square lattice and a 15-velocity cubic lattice.
We conclude with some comments on interpretation of the stability results,
entropy considerations, and extension to thermohydrodynamic LB models.

\section{Review of Chapman-Enskog Method}

This section provides a description of the Chapman-Enskog expansion applied to
the Boltzmann equation with the following definitions and conditions:

\noindent 1) The particle populations $f$ may only move with velocities that
are members of the set of discrete velocity vectors $e_{i}$. The corresponding
populations are denoted $f_{i}$.

\noindent 2) A collision operator with a single relaxation time, $\tau$, is
used to redistribute populations $f_{i}$ towards  equilibrium values
$f_{i}^{eq}$. This is also referred to as a BGK collision operator where $\tau$
is inversely proportional to density \cite{vinc}. For constant density flows
$\tau$ is a constant.

\noindent 3) The equilibrium velocity distribution function is written as a
truncated power series in the macroscopic flow velocity.

The discrete velocity Boltzmann equation then becomes
\begin{equation}
\frac{\partial f_{i}}{\partial t}
+{\bf e}_{i} \cdot {\bf \nabla}f_{i}= -\frac{1}{\tau} (f_{i}-f_{i}^{eq})
\end{equation}
where the velocity distribution function $f_{i}$ is constructed so that
macroscopic flow variables are defined by its moments:

\noindent{Mass:}
\begin{equation}
n \equiv \sum_{i}f_{i}
\end{equation}
Momentum:
\begin{equation}
n{\bf u} \equiv \sum_{i}f_{i}{\bf e}_{i}
\end{equation}

Equation (1) may be written in non-dimensional form by using a characteristic
flow length scale $L$, reference speed $e_{r}$, and density $n_{r}$. Two
reference time scales are used, $t_{c}$ to represent the time between particle
collisions and $L/e_{r}$ to represent a characteristic flow time. The reference
speed may be selected to be the magnitude of the minimum non-zero discrete
velocity. If only one speed is used, then the velocity set for the
non-dimensional equations is simply a set of unit vectors. The resulting
non-dimensional equation is
\begin{equation}
\frac{\partial \hat{f_{i}}}{\partial \hat{t}}
+ \hat{\bf e}_{i} \cdot {\hat{\bf \nabla}}\hat{f_{i}}=
-\frac{1}{\varepsilon\hat{\tau}} (\hat{f_{i}}-\hat{f_{i}}^{eq})
\end{equation}
where the caret symbol is used to denote non-dimensional quantities
$\hat{\bf e}_{i}={\bf e}_{i}/e_{r}$,$\hat{\bf \nabla}=L{\bf \nabla}$, $\hat{t}=
t e_{r}/L$,
$\hat{\tau}=\tau/t_{c}$, and $\hat{f_{i}}=f_{i}/n_{r}$. The parameter
$\varepsilon=t_{c} e_{r}/L$  and may be interpreted as either the ratio of
collision time to flow time or as the ratio of mean free path to the
characteristic flow length (i.e. Knudsen number). We will not use the caret
notation further but will assume that the equations are in non-dimensional form
henceforth.

The first step in the Chapman-Enskog procedure is to invoke a multi-scale
expansion of the time and space derivatives in the small parameter,
$\varepsilon$ as follows.
\begin{equation}
\frac{\partial}{\partial t} =
\frac{\partial}{\partial t_{1}}+
\varepsilon \frac{\partial}{\partial t_{2}}+...
\end{equation}
\begin{equation}
\nabla = \nabla_{1}+\varepsilon \nabla_{2}+...
\end{equation}

We also expand the distribution function as
\begin{equation}
f_{i} = f_{i}^{(0)}
+\varepsilon f_{i}^{(1)}  +\varepsilon^{2}f_{i}^{(2)} + ...
\end{equation}
where the zeroth-order term is the equilibrium distribution function so that
the collision operator becomes
\begin{equation}
-\frac{1}{\varepsilon\tau} (f_{i}-f_{i}^{eq})=
-\frac{1}{\tau}
( f_{i}^{(1)}  +\varepsilon f_{i}^{(2)} + ...).
\end{equation}
Since mass and momentum are conserved in collisions, the sum over the $i$
velocities of the collision term and the collision term multiplied by ${\bf
e}_{i}$ must be zero. Therefore, the sums on $f_{i}$ in equations (2) and (3)
also hold for $f_{i}^{(0)}$ and sums over nonequilibrium populations are zero.
We make the further assumption that sums over the nonequilibrium populations
corresponding to each order in $\varepsilon$ independently vanish:
$\sum_{i}f_{i}^{(l)}=0$ and $\sum_{i}{\bf e}_{i}f_{i}^{(l)}=0$ for $l>0$.

Substituting the above expansions into the Boltzmann equation, we obtain
equations of zeroth and first order in $\varepsilon$ which are written
separately as
\begin{eqnarray}
\frac{\partial}{\partial t_{1}}f^{(0)}_{i}+{\bf e}_{i} \cdot {\bf \nabla}_{1}
f^{(0)}_{i} = -\frac{1}{\tau}
f^{(1)}_{i}
\end{eqnarray}
and
\begin{eqnarray}
\frac{\partial}{\partial t_{2}}f^{(0)}_{i}
+\frac{\partial}{\partial t_{1}}f^{(1)}_{i}
+{\bf e}_{i} \cdot {\bf \nabla}_{1} f^{(1)}_{i} +{\bf e}_{i} \cdot {\bf
\nabla}_{2} f^{(0)}_{i}= -\frac{1}{\tau}
f^{(2)}_{i}
\end{eqnarray}
where it has been assumed that $\tau$ is $O(1)$.

When equations (9) and (10) are summed over the $i$ velocities the continuity
or mass conservation equation to first order in $\varepsilon$ is obtained as
\begin{eqnarray}
\frac{\partial n }{\partial t} +
{\bf \nabla}\cdot (n{\bf u}) = 0.
\end{eqnarray}
The momentum equation to first order in $\varepsilon$ is obtained by
multiplying the above equations by ${\bf e}_{i}$ and then summing over
velocities to obtain,
\begin{eqnarray}
\frac{\partial }{\partial t}(n{\bf u})
+{\bf \nabla}\cdot ({\bf \Pi}^{(0)}+ {\bf \Pi}^{(1)}) = 0
\end{eqnarray}
where ${\bf \Pi}^{(l)}$ is the momentum flux tensor and is defined as
\begin{equation}
{\bf \Pi}_{\alpha \beta}^{(l)} =
\sum_{i}e_{i \alpha}e_{i \beta}f_{i}^{(l)}
\end{equation}
for $l=0,1$.
The constitutive relations for this tensor are obtained by selecting a
particular lattice geometry and equilibrium distribution functional form and
then proceeding to match moments of the distribution function with terms in the
Navier-Stokes equations.

As an example, when this is performed for a hexagonal lattice with unit
velocity vectors defined by ${\bf e}_{i}=\{cos(2\pi (i-1)/6),sin(2\pi
(i-1)/6)\}$ for $i=$1,2,...,6, a suitable equilibrium distribution function is
found to be
\begin{equation}
f_{0}^{\rm eq} = n\alpha-n u^{2}
\end{equation}
\begin{equation}
f_{i}^{\rm eq} = \frac{n(1-\alpha)}{6}+ \frac{n}{3}{\bf e}_{i}\cdot{\bf u}
+ \frac{2n}{3}({\bf e}_{i}\cdot{\bf u})^{2}
-\frac{n}{6} u^{2}
\end{equation}
where $\alpha$ is a constant that determines the distribution of mass between
the moving and nonmoving populations \cite{hchen}.

We may readily evaluate the constitutive relations for this distribution
function by making use of the lattice relations
\begin{equation}
\sum_{i}e_{ i \alpha}e_{ i \beta} =
3\delta_{\alpha \beta}
\end{equation}

\begin{equation}
\sum_{i}e_{i \alpha}e_{ i \beta} e_{ i \gamma}
e_{ i \theta}
= \frac{3}{4} ( \delta_{\alpha \beta}\delta_{\gamma \theta}+
\delta_{\alpha \gamma}\delta_{\beta \theta}+
\delta_{\alpha \theta}\delta_{\beta \gamma}),
\end{equation}
and noting that summations of an odd number of $e_{i}$'s are equal to zero.

Substituting equation (15) into the equation (13) for ${\bf \Pi}^{(l)}$ above,
we find that
\begin{equation}
{\bf \Pi}^{(0)}_{\alpha \beta} =3 n \frac{1-\alpha}{6} \delta_{\alpha \beta}+n
u_{\alpha}u_{\beta}
\end{equation}
which gives a Galilean invariant convective term in the momentum equation. By
identifying the isotropic part of this tensor as the pressure, we obtain an
ideal gas law equation of state (i.e. $p=\frac{1-\alpha}{2}n$) and the gradient
of the pressure in the momentum equation. The other term in the momentum
equation is obtained by using Equation (9) as an expression for $f_{i}^{(1)}$
to obtain
\begin{equation}
{\bf \Pi}^{(1)}_{\alpha \beta} = -\tau\{\frac{\partial}{\partial t}{{\bf
\Pi}^{(0)}_{\alpha \beta}}+ \frac{\partial}{\partial x_{\gamma}}\sum_{i}e_{i
\alpha}e_{ i \beta} e_{ i \gamma}f_{i}^{(0)}\}.
\end{equation}
The next step in the Chapman-Enskog procedure is to replace time derivatives at
this order $\varepsilon$ level with spatial derivatives using the Euler level
equations. Thus, the time derivative of the density in the above equation may
be replaced using the continuity equation. Also, the time derivative of
$nu_{\alpha}u_{\beta}$ can be replaced using the Euler level momentum equation
which converts the time derivative to spatial derivatives as follows
\begin{equation}
\frac{\partial}{\partial t}(n u_{\alpha} u_{\beta}) =
u_{\alpha}(-\frac{\partial p}{\partial x_{\beta}}-nu_{\alpha}\frac{\partial
u_{\beta}}{\partial x_{\alpha}})+u_{\beta}(-\frac{\partial p}{\partial
x_{\alpha}}-\frac{\partial}{\partial x_{\beta}}(n u_{\alpha} u_{\beta}))
\end{equation}
where the terms of $O(u^{3})$ are neglected in the incompressible limit. The
equation of state from equation (18) is used to replace the pressure gradient
with a density gradient. Finally, when the equilibrium distribution is
substituted into the last term of equation (19), the only term  that remains is
the ${\bf e}_{i} \cdot {\bf u}$ term which is evaluated using Equation (17).

Upon substitution into equation (12), the final form of the momentum equation
is
\begin{equation}
n\frac{\partial u_{\alpha}}{\partial t}+nu_{\beta}\frac{\partial
u_{\alpha}}{\partial x_{\beta}}=
-\frac{\partial p}{\partial x_{\alpha}}+\frac{\partial}
{\partial x_{\beta}}(\frac{\lambda}{n}(\frac{\partial nu_{\gamma}}
{\partial x_{\gamma}}+u_{\alpha}\frac{\partial n}{\partial
x_{\beta}}+u_{\beta}\frac{\partial n}{\partial x_{\alpha}}))+\frac{\partial}
{\partial x_{\beta}}(\mu(\frac{\partial u_{\beta}}
{\partial x_{\alpha}}+\frac{\partial u_{\alpha}}
{\partial x_{\beta}}))
\end{equation}
where
\begin{equation}
\mu=\frac{\tau n}{4}
\end{equation}
and
\begin{equation}
\lambda=\frac{\tau n(2\alpha-1)}{4}.
\end{equation}
In two dimensions, the bulk viscosity is the sum of these two so that
\begin{equation}
K=\frac{\tau n\alpha}{2}
\end{equation}
which gives zero bulk viscosity as expected for the monatomic gas when energy
is conserved (i.e. when $\alpha=0$ it can be shown that conservation of mass is
equivalent to conservation of energy).

Note that these equations are not the standard Navier-Stokes equations because
there are derivatives of the density in the second viscosity term on the right
side of the equation. If these gradients of density are negligible this
hexagonal lattice, discrete Boltzmann equation should behave approximately as
the Navier-Stokes equations. Since the gradients of the density are $O(u^{2})$
(see references \cite{mart} and \cite{majda}), the unphysical terms in equation
(21) are $O(u^{3})$. Thus, although the physics contains compressibility
effects (that differ from the compressible Navier-Stokes equations), one may
come arbitrarily close to solving incompressible flow by reducing the Mach
number and thereby allowing information to propagate throughout the domain
while little convection occurs. For this reason, no Poisson solver is required
to determine the pressure and simple particle reflections at boundaries may be
used to invoke no-slip conditions. We also note that if the second viscosity
$\lambda$ is zero, the complete compressible Navier-Stokes equations are given
but the bulk viscosity is then nonzero.

There are differences between the incompressible Navier-Stokes equations and
the macroscopic behavior of the discrete-velocity Boltzmann equations because
of the asymptotic nature of the Chapman-Enskog method. The differences may be
attributed to Burnett level and higher level terms or as small deviations from
the above relation for the kinematic viscosity. For this reason, previous LB
studies have reported comparisons between the Chapman-Enskog prediction and
numerical simulation measurements of the viscosity (e.g. Kadanoff {\em et. al.}
\cite{kad}). However, the Burnett level terms are expected to become negligible
as the global Knudsen number becomes small. Since the Knudsen number is
proportional to the Mach number divided by the Reynolds number, the Burnett
terms may be classified with other ``compressibility" effects and should become
small as the Mach number approaches zero for a fixed Reynolds number.

In conclusion, the discrete Boltzmann equation in dimensionless form, equation
(4), may be discretized and numerically simulated to provide approximate
solution to the continuity and momentum equations given by equations (11) and
(21), respectively. The results can then be put back into dimensional form
using the reference quantities. Simulations may come arbitrarily close to
incompressible Navier-Stokes behavior with differences being attributed solely
to discretization and compressibility effects.

\section{The Lattice Boltzmann Discretization}

At this point we will narrow our view to a particular discretization of the
non-dimensional discrete Boltzmann equation. In particular, we will choose the
lattice-Boltzmann method which is an exact Lagrangian solution for the
convective derivatives. For a given convection velocity, this type of scheme is
typically obtained by using an Euler time step in conjunction with an upwind
spatial discretization and then setting the grid spacing divided by the time
step equal to the velocity. Discretization of equation (4) results in the
following equation.
\begin{equation}
\frac{f_{i}({\bf x},t+\Delta t)-f_{i}({\bf x},t)}{\Delta t}+ \frac {f_{i}({\bf
x} + {\bf e}_i \Delta x,t +\Delta t)-f_{i}({\bf x},t+\Delta t)}{\Delta
x}=-\frac{(f_{i}({\bf x},t)-f_{i}^{(0)}({\bf x},t))}{\epsilon \tau}.
\end{equation}

Lagrangian behavior is then obtained by the selection of the lattice spacing
divided by the time step to equal the magnitude of ${\bf e}_{i}$, which was
normalized so that the smallest velocity magnitude is unity. When the equation
is multiplied by $\Delta t$, the result is the cancellation of two terms on the
left side of the above equation leaving only one term evaluated at $t+\Delta t$
so that the method is explicit.

The next characteristic of the lattice Boltzmann method is the selection of the
time step to equal the reference collision time. The result is the cancellation
of the Knudsen number in the denominator of the collision term giving the
following simple form that is commonly referred to as the lattice Boltzmann
equation (LBE).
\begin{equation}
f_{i}({\bf x} + {\bf e}_{i}\Delta t,t + \Delta t)-f_{i}({\bf x},t)=
-\frac{1}{\tau}(f_{i}({\bf x},t)-f_{i}^{(0)}({\bf x},t)).
\end{equation}

This equation has a particularly simple physical interpretation in which the
collision term is evaluated locally and there is only one streaming step or
``shift" operation per lattice velocity. This stream-and-collide particle
interpretation is a result of the fully-Lagrangian character of the equation
for which the lattice spacing is the distance travelled by the particles during
a time step. Higher order discretizations of the discrete Boltzmann equation
typically require several ``shift" operations for the evaluation of each
derivative and a particle interpretation is less obvious. In fact, the entire
derivation of the LB method was originally based on the idea of generalizing LG
models by solving the LG Boltzmann equation and relaxing the exclusion
principle that particle populations be either zero or one for each velocity
\cite{mcn}. It did not originally occur to the authors that the LB method could
be considered a particular discretization for the discrete Boltzmann equation
\cite{guy1}.

The particle model allows boundary conditions to be implemented as particular
types of collisions. If populations are reflected directly back along the
lattice vector along which they streamed, the result is a ``no-slip" velocity
boundary condition. One may also define specular reflection conditions that
yield a slip condition. Models for which energy is conserved allow
specification of heat-transfer boundary conditions using particle reflection
conditions as well \cite{alex}. These simple boundary conditions make the LB
method particularly suited to parallel computing environments and the
simulation of flows in complex geometries.

Although first order discretizations have been used, the LB method is typically
considered to be a second order method because contributions that result from
discretization error are taken to represent physics \cite{mario}. The inclusion
of numerical viscosity is accomplished by Taylor expanding equation (26) about
$x$ and $t$. When the second order terms in this expansion are included in the
above Chapman-Enskog analysis, the result is that the coefficient $\tau$ in the
transport coefficients is simply replaced by $\tau-\frac{1}{2}$ (see reference
\cite{alex}). Thus, the lattice contribution to the viscosity for this LB
scheme is negative, requiring the value of the relaxation time to be greater
than half of the time step to maintain positive viscosity. Note that
third-order terms in the Taylor-series expansion are necessarily of order
$\epsilon^3$ in the Chapman-Enskog expansion. Thus, as with traditional kinetic
theory, there may be some error arising from the Burnett level terms.

Since the LB method under consideration is valid only in the incompressible
limit, the main dimensionless parameter of interest is the Reynolds number.
Convergence of the solution to the incompressible Navier-Stokes equations for a
fixed Reynolds number is then obtained by letting the Mach number become small
enough to remove compressibility effects, and letting the lattice spacing ${\bf
e}_{i} \Delta t$ become small enough to ``resolve" the flow.  Reverting to the
caret notation for dimensionless quantities, the Reynold's number for the
hexagonal lattice may now be written
\begin{equation}
Re=\frac{LU}{\nu}=\frac{4 N \hat{U}}{\hat{\tau}-\frac{1}{2}}.
\end{equation}
where $N=\frac{L}{\Delta x}$ is the number of lattice spaces. The dimensionless
velocity is the characteristic Mach number which should be small to simulate
incompressible flow. Thus, the convergence at a given Reynolds number is
performed by increasing $N$ while either increasing $\hat{\tau}$ and/or
decreasing $\hat{U}$ appropriately. For a  decrease in the value of $\hat{U}$,
a proportionate increase in the number of time steps is needed to reach the
same flow evolution time.

Concluding, the LB method makes use of first order discretizations of the
dimensionless discrete velocity Boltzmann equation in both time and space. The
dimensionless time step and lattice spacing are set equal and numerical
contributions to viscosity are accounted for and considered to be part of the
physics of the method. With these effects included, the LB method is a second
order method in both space and time for the simulation of the Navier-Stokes
equations. In the use of LB models developed for incompressible Navier-Stokes
simulation care must be taken to ensure that both the Mach number and the
Knudsen number are small enough that the deviation from incompressible behavior
is negligible.

\section{Lattice Boltzmann Linear Stability}
\subsection{Theory Development}
The lattice Boltzmann equation, equation (26), is an explicit scheme for the
computation of the particle population associated with each discrete velocity.
It is a nonlinear scheme due to the use of the equilibrium distribution
function in the collision term. This function is quadratic in velocity (cubic
for energy conserving models) and the density and velocity are computed as sums
over all of the populations at a site. In an effort to assess the numerical
stability of LB schemes with a linearized collision operator, Benzi $\em et.
al.$ \cite{benzi} and Grunau \cite{daryl} performed stability analyses by
neglecting nonlinear terms. Their linear analysis is equivalent to studying the
stability under conditions of no mean flow or the stability of a finite mean
flow that is uniform in space (i.e. zero wavenumber perturbations). However, a
von Neumann linearized stability analysis of the LB scheme requires the
linearization of all nonlinear terms about global equilibrium values of the
populations (denoted by the overbar) that are based on some mean density,
velocity, and internal energy for energy-conserving models. Thus, we expand
$f_{i}$ as
\begin{equation}
f_{i}(x,t)=\overline{f_{i}^{(0)}}+f_{i}'(x,t)
\end{equation}
where the equilibrium populations $\overline{f_{i}^{(0)}}$ are constants that
do not vary in space or time and depend only on the mean density and velocity.
The fluctuating quantities $f_{i}'$ are not equal to $f_{i}^{(1)}$ because we
have linearized about the equilibrium populations evaluated for a mean density
and mean velocity. However, the density and velocity deviate from the mean
values such that the equilibrium populations vary in space and time. If the
perturbations are uniform in space, $f_{i}'=f_{i}^{(1)}$ and we recover the
stability results for the collision term.

We define the update operator for populations $f_{i}$ to be \begin{equation}
g_{i}(f_{j}) =f_{i}({\bf x},t)
-\frac{1}{\tau}(f_{i}({\bf x},t)-f_{i}^{(0)}({\bf x},t))
\end{equation}
where all of the $j$ populations at a site enter through the equilibrium
distribution function on the right side of the equation.
Taylor expanding $g$ about $\overline{f_{i}^{(0)}}$ results in the following
equation.
\begin{equation}
\overline{f_{i}^{(0)}}+f_{i}'({\bf x} + {\bf e}_{i}\Delta t,t +\Delta
t)=g_{i}(\overline{f_{j}^{(0)}})+ \frac{\partial
g_{i}(\overline{f_{j}^{(0)}})}{\partial f_{j}}f_{j}'(x,t)+O(f_{j}'(x,t)^2)
\end{equation}
Since $\overline{f_{i}^{(0)}}=g_{i}(\overline{f_{j}^{(0)}})$, the resulting
linearized system is
\begin{equation}
f_{i}'({\bf x} + {\bf e}_{i}\Delta t,t +\Delta t)=G_{ij}f_{j}'(x,t),
\end{equation}
where $G_{ij}$ is the Jacobian matrix corresponding to the coefficient of the
linear term in equation (30) and does not depend on location or time.

Spatial dependence of the stability is investigated by taking the Fourier
transform of equation (31) to obtain
\begin{equation}
F_{i}({\bf k},t+\Delta t)=\Gamma_{ij} F_{i}({\bf k},t)
\end{equation}
where
\begin{equation}
\Gamma_{ij}= diag\{{\em exp}(-{\em i}{\bf k} \cdot {\bf e}_{j})\} G_{ij}
\end{equation}
and the wavenumber has units of inverse lattice spacing. These units are not
the most common form for presentation: if we define wavenumber using ${\em
exp}(-2\pi{\em i}{\bf k} \cdot {\bf e}_{j}\Delta t)$, then $k$ is the number of
sine waves in the domain and the highest resolution wavenumber is $1/(2\Delta
x)$.

We observe that if the wavenumber is zero, the first matrix becomes the
identity matrix and the eigenvalues of $G_{ij}$ determine stability. In this
case of uniform flow, if the eigenvalues of $G_{ij}$ have modulus less than
unity, then the scheme is asymptotically stable. The eigenvalues are
$\{1,1-\frac{1}{\tau}\}$ where the unity eigenvalues have multiplicity $D+1$ in
$D$ dimensions, corresponding to microscopic mass and momentum conservation.
Thus, stability of uniform flows is guaranteed if $\tau > \frac{1}{2}$.

The elements of the matrix $G_{ij}$ include the the linearization of the
nonlinear terms in the equilibrium distribution function. As an example, the
derivative with respect to $f_{j}$ of the first nonlinear term of the
equilibrium distribution function, $n({\bf e}_{i} \cdot {\bf u})^2$ is
\begin{equation}
2({\bf e}_{i} \cdot {\bf e}_{j})({\bf e}_{i} \cdot {\bar {\bf u}})-({\bf e}_{i}
\cdot {\bar {\bf u}})^2.
\end{equation}

Stability has been investigated by using $Mathematica^{TM}$ version 1.2 to
solve for eigenvalues of $\Gamma_{ij}$ both algebraically and numerically for
several lattices and associated equilibria. The following sections document the
stability boundaries as functions of the following five parameters; wavenumber
${\bf k}$, relaxation parameter $\tau$, velocity $\bar {\bf u}$, and particle
population distribution parameters $\alpha$ and $\beta$ (introduced for square
and cubic lattices below).

\subsection{7-Velocity Hexagonal Lattice Results}

The lattice definition and equilibrium velocity distribution function for the
hexagonal lattice is described in Section 2 above. When the lattice Boltzmann
equation is linearized about a mean velocity and density, and a Fourier
transform is performed, the eigenvalues of the resulting Jacobian matrix
$\Gamma_{ij}$ may be evaluated to assess linear stability of the system. As
mentioned above, if the wavenumber is zero, $\tau = \frac{1}{2}$ is the only
linear stability boundary. Indeed, numerical simulation results are
consistently unstable if the value of $\tau$ is too close to 0.5.
This boundary has been well tested because there is considerable interest in
using this LB method to simulate high-Reynolds number flow and as $\tau$
appoaches 0.5, the Reynolds number approaches infinity. The fact that values of
$\tau$ slightly greater than one-half can lead to instability is attributed to
nonlinear effects.

The linearized stability in the hexagonal case depends on the four parameters
$\tau$,$\alpha$, $\bar {\bf u}$, and $\bf k$. Therefore, a complete mapping of
all stability boundaries is not computationally feasible for even this
7-velocity model. Since the velocity and wavenumber are both vectors, a study
was performed in which the angle between these vectors was varied while the
other parameters remained fixed. The result for the case studied was that the
most unstable condition occured when the angle between the vectors was equal to
zero. Although there is no proof that this result holds for all parameter
values, we have assumed that the velocity and the wavenumber vectors are
aligned with the first velocity vector for each lattice (i.e. the horizontal
axis). This assumption was made for all of the results that follow.

The second attempt at simplifying the analysis was to determine if there was a
single wavenumber that was consistently the most unstable. When using a unit
lattice spacing, the highest resolvable wavenumber is equal to $\pi$. Figure 1
is a plot of the maximum eigenvalue magnitude of $\Gamma_{ij}$ as a function of
wavenumber for two unstable conditions when $\tau=.5$. The solid line
corresponds to $\bar {\bf u}=.2$ and $\alpha=.2$ and the dotted line is for
$\bar {\bf u}=.23$ and $\alpha=.3$. There are two unstable eigenvalues in the
first case and only one in the second case. It is clear that the most unstable
wavenumber changes from around $\pi$ to a value less than 2.0 so that there is
not a single wavenumber that is always the most unstable. Therefore subsequent
studies  evaluate eigenvalues at wavenumbers from 0.1 to 3.1 in steps of 0.2
and the wavenumber with the largest eigenvalue modulus is considered to be the
``most unstable wavenumber". The coarse wavenumber resolution undoubtedly
results in stability boundaries that are actually in an unstable parameter
range. In other words, stability boundaries in the following results should be
shifted slightly towards the stable parameter domain.

The distribution of the mass between the non-moving population and the six
moving populations is controlled by the parameter $\alpha$. Since we are
usually interested in high-Reynolds number flows, we investigated the stability
of the method as a function of $\alpha$ and $\bar {\bf u}$ when
$\tau=\frac{1}{2}$. An iterative scheme was used in which a value of $\alpha$
was selected and $\bar {\bf u}$ was incrementally increased until the maximum
eigenvalue modulus exceeded unity. The neutral stability boundary was obtained
in this manner by varying the value of $\alpha$ from zero (equivalent to an
energy-conserving model) to near unity (for which almost all of the mass is
stationary). The resulting boundary is plotted as the left curve in Figure 2.
As the value of $\alpha$ is increased from near zero, the velocity for which
the LB scheme is stable increases to a maximum of around one-third when
$\alpha$ is near two-thirds. As $\alpha$ increases further, however, the
maximum stable velocity again decreases. The data in Figure 1 were taken near
the kink in the stability boundary curve near $\bar {\bf u}=.2$. The kinks are
caused when a parameter change results in the most unstable wavenumber shifting
to a different eigenvalue.

It is well known from simulations that as $\tau$ is increased, the LB method
becomes stable at higher values of velocity for a given value of $\alpha$. A
study of this effect was performed by selecting $\alpha=0.7$ (near the most
stable value in Figure 2) and then iterating the mean flow velocity and the
relaxation time to determine the neutral stability boundary. The results are
presented in Figure 3 as the solid curve. When $\tau=\frac{1}{2}$ we see in
both Figure 2 and Figure 3 that the maximum stable mean flow velocity is near
0.32. As the value of $\tau$ increases, the maximum stable velocity is seen to
decrease slightly and then increase and level off at a value of about 0.39.
Thus, the maximum velocity should be small to 1) retain a stable scheme and 2)
keep higher-order terms from the Chapman-Enskog expansion negligible. Since the
velocity is limited, high Reynolds number flow is obtained by either increasing
resolution or decreasing $\tau$ to values near one-half (i.e. near the linear
stability boundary).

\subsection{9-Velocity Square Lattice Results}

Another lattice that is commonly used for two-dimensional incompressible flow
simulations is the 9-velocity square lattice defined by vectors, ${\bf
e}_{i}^{I}=\{cos(\pi(i-1)/2),sin(\pi(i-1)/2)\}$ and ${\bf
e}_{i}^{II}=\{cos(\pi(i-\frac{1}{2})/2),sin(\pi(i-\frac{1}{2})/2)\}$ for
$i=1,4$.
The equilibrium distribution function for these moving populations and a
non-moving population is given by
\begin{equation}
f_{0}^{\rm eq} = n\alpha-\frac{2}{3}n u^{2}
\end{equation}
\begin{equation}
f_{i}^{I,\rm eq} = n\beta+ \frac{n}{3}{\bf e}_{i}^{I}\cdot{\bf u}
+ \frac{n}{2}({\bf e}_{i}^{I}\cdot{\bf u})^{2}
-\frac{n}{6} u^{2}
\end{equation}
\begin{equation}
f_{i}^{II,\rm eq} = n\frac{(1-4\beta-\alpha)}{4}+ \frac{n}{12}{\bf
e}_{i}^{II}\cdot{\bf u}
+ \frac{n}{8}({\bf e}_{i}^{II}\cdot{\bf u})^{2}
-\frac{n}{24} u^{2}.
\end{equation}
The Jacobian matrix $\Gamma_{ij}$ for this system is a 9x9 matrix which again
gives $\tau=\frac{1}{2}$ as the only stability boundary for homogeneous flow
(${\bf k}=0$). The first numerical study for this lattice was to determine if
the most unstable wavenumber occurred at a single value. Unlike the results
shown in Figure 1 for the hexagonal lattice, the most unstable wavenumber was
consistently equal to about 2.3 when $\tau=\frac{1}{2}$. Thus, the following
studies for which $\tau=\frac{1}{2}$ did not require evaluation at many
wavenumbers but simply used this most unstable wavenumber.

With both $\alpha$ and $\beta$ as mass distribution parameters, there are five
parameters in the matrix $\Gamma_{ij}$. The next numerical study performed on
this system addressed the stability for various mass distributions for fixed
mean speed and relaxation time for the most unstable wavenumber. The dotted
lines in Figure 4 delineate the stability boundaries when $\tau=.5$ and ${\bf
u}=.3$. Combinations of $\alpha$ and $\beta$ that lie between the two dotted
lines result in linear stability while combinations to the left and right of
the dotted lines result in linear instability. The right dotted line is
parallel to the curve We note that the values of $\alpha=\frac{4}{9}$ and
$\beta=\frac{1}{9}$ used in reference \cite{mart} lie in the stable domain.
Also, these particular values cause the second viscosity to be identically zero
($\lambda=0$) so that compressible Navier-Stokes equations (11) and (21) are
recovered but the bulk viscosity is equal to the shear viscosity.

The strip of stable eigenvalues in Figure 4 allows us to eliminate the
parameter $\beta$ for subsequent parameter studies by enforcing a parametric
relation with $\alpha$. We originally chose
$\beta=\frac{1}{4}-\frac{\alpha}{3}$ to fall within the stable strip of values.
However, this relation plotted as a line on Figure 4 would lie in the center of
the stable range when $\beta=0$ and $\alpha=\frac{3}{4}$ but when
$\beta=\frac{1}{4}$ and $\alpha=0$ the line would fall just outside the stable
range. Nonetheless, this relation was used in the following studies.

The stability boundary for the square lattice is plotted as the right curve in
Figure 2 when the above parametric relation for $\beta$ is used,
$\tau=\frac{1}{2}$, and the most unstable wavenumber is used. From Figure 4,
the parametric relation for $\beta$ indicates instability for ${\bf u}=.3$ when
$\alpha=0$. This result can also be seen in Figure 2 which shows that the
neutral stability boundary at $\alpha=0$ occurs for the velocity just under
$0.3$. The most interesting result seen in this figure however, is that for
values of $\alpha$ greater than about $0.2$, the maximum stable velocity is a
constant near $\frac{1}{3}$. We have not been able to identify an analytic
reason that $\bar {\bf u}=\frac{1}{3}$ is the stability boundary and is
independent of $\alpha$ in the center of the stable parameter strip seen in
Figure 4.

Since the stability is independent of $\alpha$ over a wide range of values, we
have used the values of reference \cite{mart} to study the stability
characteristics as the relaxation time is varied. The neutral stability
boundary is plotted as the middle curve (dot-dash) in Figure 3 when
$\alpha=\frac{4}{9}$, $\beta=\frac{1}{9}$, and the most unstable wave number
(not necessarily 2.3 when $\tau$ varies) is considered. The results are similar
to the solid curve in Figure 3 which was discussed in the hexagonal lattice
results.
As $\tau$ is increased from one-half, the maximum stable mean flow velocity
increases monotonically from about one-third to a value near 0.42 when $\tau$
is near 0.68. However, the maximum stable velocity does not change for further
increases of $\tau$. The kink in the curve is a result of the shift of the most
unstable eigenvalue/wavenumber to another eigenvalue for which the most
unstable wavenumber is $\pi/2$. High Reynolds number flow is obtained by
allowing $\tau$ to approach the stability boundary of $\tau=\frac{1}{2}$, in
accordance with the results for the hexagonal lattice discussed above.

\subsection{15-Velocity Cubic Lattice Results}
A simple way to extend the square lattice, with vectors to the sides and
corners of the square, to three dimensions is to use vectors to the sides and
corners of a cube \cite{jane}\cite{qian}\cite{alex2}. This defines a
body-centered-cubic lattice with ${\bf e}_ {i}^{I} \in ({\pm 1},0,0),
(0,{\pm 1},0)$, $(0,0,{\pm 1})$ and ${\bf e}_ {i}^{II} \in
({\pm 1}, {\pm 1}, {\pm 1})$.
The equilibrium distribution function for these moving populations and a
non-moving population is given by
\begin{eqnarray}
f^{\rm (eq)}_{0} = \alpha n- \frac{n}{3}{\bf u}^2,
\end{eqnarray}
\begin{eqnarray}
f^{I,{\rm (eq)}}_{i} = \beta n
+ \frac{n}{3}({\bf e}_{ i}^{I} \cdot {\bf u})
+\frac{n}{2}({\bf e}_{i}^{I} \cdot {\bf u})^{2} -  \frac{n}{6} {\bf u}^{2},
\end{eqnarray}
for ${\bf e}_{i}^{I}$ along the lattice axes, and
\begin{eqnarray}
f^{II,{\rm (eq)}}_{i} = \frac{(1-6\beta-\alpha)}{8} n
+ \frac{n}{24}({\bf e}_{ i}^{II} \cdot {\bf u})
+\frac{n}{16}({\bf e}_{i}^{II} \cdot {\bf u})^{2} -  \frac{n}{48} {\bf u}^{2}
\end{eqnarray}
for ${\bf e}_{i}^{II}$ along the links to the corners of the cube.

As in the case of the 9-velocity 2-D model, the first numerical study performed
on this system addressed the determination of the most unstable wavenumber.
Because of the similarities in the lattice definitions, the most unstable
eigenvalue again occurs at wavenumber equal to 2.3 for unit lattice spacing
when $\tau=\frac{1}{2}$.

Following the investigation discussed above for the 2-D square lattice, the
next investigation studied the dependence of stability on the mass distribution
parameters $\alpha$ and $\beta$ for fixed mean speed and relaxation time for
the most unstable wavenumber. The solid lines in Figure 4 delineate the
stability boundaries when $\tau=.5$ and ${\bar{\bf u}}=.32$. Combinations of
$\alpha$ and $\beta$ that lie between the two lines result in linear stability
while combinations to the left and right of the dotted lines result in linear
instability. Values of $\alpha=\frac{1}{8}$ and $\beta=\frac{1}{8}$ used in
reference \cite{alex2} lie near the top and left of the stable domain seen in
Figure 4.

A parametric relation between $\alpha$ and $\beta$ was chosen to fall along the
strip of stable values from Figure 4. The relation was $\beta=0.2-0.3\alpha$
which lies near the center of the strip for all values (in contrast with the
square lattice relation that fell just outside the stable strip for small
$\alpha$ values). Using this relation, the following results were similar to
those found in the case of the 9-velocity square lattice.

For $\tau=0.5$, and ${\bf k}=2.3$, the linear stability boundary was computed
for varying $\alpha$ and $\bar{\bf u}$. As in the case of the 2-D square
lattice, the neutral stability boundary was found to occur for a mean velocity
of about one-third independent of $\alpha$. A plot of this curve would appear
as a vertical line on top of the square lattice line in Figure 2. Finally, a
stability boundary was found for $\alpha=\frac{1}{8}$ and the most unstable
wavenumber for varying $\tau$ and $\bar{\bf u}$. The resulting stability
boundary is plotted as the right curve (dashed) in Figure 3 verifying that the
cubic lattice stability results are  very similar to the square lattice results
when the mass distribution parameters are selected as discussed above. The main
difference is that the cubic lattice has a larger maximum stable mean flow
velocity that is near 0.475 for $\tau$ above about 0.7.

\section{Conclusions}
The lattice Boltzmann equation is viewed as a Lagrangian finite-difference
numerical approximation to the discrete-velocity Boltzmann equation that makes
use of a BGK collision operator. The collision serves to relax the velocity
distribution function towards an equilibrium distribution that is selected so
that the first few velocity moments match those of the Maxwell-Boltzmann
distribution. Thus, models have been developed for which a Chapman-Enskog
expansion predicts second-order numerical accuracy for the solution of the
incompressible Navier-Stokes equations. In addition to conserving mass and
momentum during collision, the aforementioned matching criteria are also
required, with the result that entropy is not necessarily increased during the
collision. As a finite difference scheme that does not provide an H-theorem for
the particle model, it is not surprising that numerical instability can and
frequently does arise during simulation. For this reason, a linearized
stability analysis was performed on the hexagonal, square, and cubic lattices
defined above.

Linearization of the population $f_{i}$ is performed about an equlibrium value
that does not vary in space or time and depends only on mean density and
velocity. We then investigate whether perturbations in the populations grow or
decay. The linear stability of the LB models depends on the mass distribution
parameters, the mean velocity, the relaxation time, and the wavenumber. The
matrix sizes are too large for the analysis to cover all of the parameter
space. Thus, numerical evaluation of the eigenvalues of the Jacobian stability
matrix was performed for various parameter values to gain some understanding of
the stability characteristics.

The main stability boundaries common to all three lattices are the following.

\noindent 1)  A well-known stability boundary requires that the relaxation time
be greater than one-half. Note that $\tau=\frac{1}{2}$ corresponds to zero
shear viscosity. Since we are often interested in high Reynolds number flows,
analysis is commonly performed along the stability boundary $\tau=\frac{1}{2}$.

\noindent 2) Another stability boundary requires the mean flow velocity to be
below a maximum stable velocity that is a function of the other parameters.

\noindent 3) As $\tau$ is increase from one-half, the maximum stable velocity
increases monotonically until a limit is reached. For the cases studied, the
limit was around 0.39, 0.42,and 0.47 for the hexagonal, square and cubic
lattices, respectively.

These boundaries require all eigenvalues have an absolute value less than or
equal to unity for all wavenumbers. Thus, numerical determination of the
stability boundaries requires the determination of the most unstable
wavenumber. As parameters are varied for the hexagonal lattice, the wavenumber
that has the largest eigenvalue modulus changes considerably. Therefore,
analysis was performed by sweeping through the entire range of wavenumbers
while varying the other parameters. However, for the square and cubic lattices,
the most unstable wavenumber was equal to 2.3 for values of $\tau$ near
one-half.

One of the main results from this study is that for the hexagonal lattice there
is a most stable value of the mass distribution parameter $\alpha=\frac{2}{3}$
which places two-thirds of the mass in the non-moving population. For this
value of $\alpha$, the relaxation time $\tau$ was increased from one-half with
the result that the maximum stable velocity increases monotonically with an
asymptote for large $\tau$ around ${\bar{\bf u}}=.39$.

Both the square and cubic lattices provide stable behavior only when the values
of the mass distribution parameters fall within certain ranges. A parametric
relation between $\alpha$ and $\beta$ can be selected which is consistenly
stable (for the mean flow velocity below some fixed value when
$\tau>\frac{1}{2}$). Using this parametric relation, an important result of
this study is that the maximum stable velocity is independent of $\alpha$ and
hence $\beta$ for a fixed $\tau$. As $\tau$ is increased, as with the hexagonal
lattice, the maximum stable velocity monotonically increases. However, when
$\tau$ reaches some critical value, the most unstable wavenumber switches to a
new eigenvalue that provides an upper limit on the maximum stable velocity
equal to about 0.42 and 0.47 for the square and cubic lattices, respectively.

These results provide some stability guidelines for researchers using LB
methods. Simulations performed too near the stability boundaries have been
observed to go unstable. A common manifestation of instability is that as a
given flow evolves, localized regions develop large velocities and instability
ensues. We note that parameters resulting in stable flow consistently provide
flow speeds and speeds of sound less than the lattice spacing divided by the
time step. For this reason, a Courant stability condition is superceded by a
more stringent stability condition on the speeds. Another stability boundary
common for finite-difference methods requires that the viscous diffusion speed
be less than the lattice spacing divided by the time step. This boundary is not
observed for LB methods because as the viscosity increases, errors of the
scheme increase due to the presence of large nonequilibrium populations, but
stability is still maintained.

As indicated in Section 2, the accuracy of the method for simulating the
incompressible Navier-Stokes equations is expected to improve as the number of
lattice sites is increasesd and as the Mach number is decreased. Quantification
of these accuracy issues is presented in reference \cite{reider}. The result is
that for models valid only in the incompressible-limit, the velocity should be
small for both stability and accuracy. Note however that the time required for
significant flow evolution (eddy-turnover time) is inversely proportional to
the velocity so that one should select the maximum velocity that is both stable
and provides compressibility errors within some desired level.

\section*{Acknowledgments}
We thank F. J. Alexander, M. G. Ancona, G. D. Doolen, D. W. Grunau, S. Hou, D.
O. Martinez, W. H. Matthaeus and M.  B. Reider for discussions and helpful
suggestions. The work was supported by U.S. Department of Energy at Los Alamos
National Laboratory.  J.D.S. thanks G. Doolen, J. Rodgers, and the Center for
Nonlinear Studies for sponsoring his stay in Los Alamos.

\newpage

\section*{Figure Captions}
\noindent
Fig.1 : Hexagonal lattice maximum eigenvalue magnitude as function of
wavenumber for $\tau=0.5$ for two cases. Dashed line is for $\alpha=.3$ and
${\bar{\bf u}}=.23$. Solid line is for $\alpha=.2$ and ${\bar{\bf u}}=.2$.

\noindent
Fig.2 :  Stability boundaries as function of ${\bar{\bf u}}$ and $\alpha$ for
most unstable wavenumber for $\tau=0.5$. The left curve is the neutral
stability boundary for the hexagonal lattice. The right curve is the neutral
stability curve for the square lattice for mass distribution parameters related
by $\beta=\frac{1}{4}-\frac{\alpha}{3}$.

\noindent
Fig.3 :  Stability boundaries as function of ${\bar{\bf u}}$ and $\tau$ for
most unstable wavenumber. The solid line is the neutral stability curve for the
hexagonal lattice for $\alpha=.7$. The middle curve (dash-dot) is the neutral
stability curve for the square lattice for $\alpha=\frac{4}{9}$ and
$\beta=\frac{1}{9}$. The dashed curve is the neutral stability curve for the
cubic lattice for $\alpha=\frac{1}{8}$ and $\beta=0.1625$.

\noindent
Fig. 4 :  Stability boundaries as function of mass distribution parameters for
most unstable wavenumber for $\tau=0.5$. The region between the two dotted
lines is the stable range for the square lattice for ${\bar {\bf u}}=0.3$. The
region between the two solid lines is the stable range for the cubic lattice
for ${\bar {\bf u}}=0.32$ .

\end{document}